\documentclass[prl,twocolumn,showpacs,a4]{revtex4}
\usepackage{amsmath}
\usepackage{bm}
\usepackage{graphicx}
\usepackage{graphpap}

\def\Journal#1#2#3#4#5#6{#5 {#1} {\bf #2} #3}

\def\CQG{\em Class. Quantum Grav.}

\def\PRD{\em Phys. Rev. D }
\def\GRG{\em Gen. Rel. Grav.}

\def\PRL{\em Phys. Rev. Lett.}

\def\PLB{\em Phys. Lett. B } 

\def\S{\Sigma}

\def\doo{d \Omega^2_k}

\def\lambdafour{\Lambda_{4}}

\def\aa{\bar a}

\def\Br{\tilde{r}}

\def\Blambda{\tilde{\lambda}}
\def\Bt{{\tilde{t}\,}}

\def\ttt{\tau}

\def\ff{N}

\begin{document} 

\title{Is the accelerated expansion evidence of a forthcoming change of signature on the brane?
}
\author{Marc Mars}
\affiliation{Dept. de F\'{\i}sica Fundamental, Universidad de Salamanca,
Plaza de la Merced s/n, 37008 Salamanca, Spain}
\author{Jos\'e M. M. Senovilla}
\author{Ra\"ul Vera}
\affiliation{F\'{\i}sica Te\'orica,  
Universidad del Pa\'{\i}s Vasco, Apartado 644, 48080 Bilbao, Spain}


\begin{abstract}
We show that regular {\em changes of signature} on brane-worlds in
AdS$_{5}$ bulks may account for some types of the recently fashionable sudden
singularities. Therefore, the possibility that the Universe seems to approach
a future sudden singularity at an accelerated rate of expansion might
simply be an indication that our braneworld is about to change from
Lorentzian to Euclidean signature. Both the brane and the bulk remain
fully regular everywhere. We present a model in which
the weak and strong energy conditions hold on the brane,
in contrast with the standard cosmologies leading to
the analogous kinematical behaviour (that is,
with a diverging Hubble factor).
\end{abstract}


\maketitle
 
Recently \cite{letter,GI,signlong}, the idea that brane-world models
\cite{branes,roy_review} are a natural scenario for the
{\em regular} description of a change of the spacetime signature
has been advocated. One of the interesting, and intriguing,
properties of these signature-changing branes is that, even though the
change of signature may be conceived as a dramatical event {\em within
the brane}, both the bulk and the brane can be fully smooth.  In
particular, observers living in the brane {\em but} assuming that
their Universe is Lorentzian everywhere may be misled to {\em
interpret} that a curvature singularity arises precisely at
the signature change \cite{letter,signlong}.

In this note we show that a correct description of this
misinterpretation might explain an accelerated expansion of the
Universe, while keeping the energy density and the rest of physical
variables regular and non-negative everywhere; in particular
without violating the
weak or strong energy conditions. Explicit models are built where
the Hubble parameter certainly increases and eventually
diverges (for pure Lorentzian branes with finite Hubble factor see \cite{SS}), but such that this corresponds to a pathology of the proper
time, which is about to disappear mutating into a spacelike
coordinate at the change of signature. The resulting accelerated expansion epoch does not need of dark, phantom, or any other exotic energy.
 
For our purposes here we only need to consider anti de
Sitter (AdS${}_{5}$) bulks.  Suitable coordinate systems allow us to
write the bulk line-element as
\[
ds^2=-(k+\lambda^2 r^2)dt^2+(k+\lambda^2 r^2)^{-1}d r^2+ r^2\doo,
\]
where $\lambda$ is a positive constant
related to the negative cosmological constant of the spacetime by
$\Lambda_5 =-6 \lambda^2$, and 
$\doo$ 
is the $3$-dimensional
metric of constant sectional curvature $k=1,0,-1$.
When $k=0,1$ the ranges of the
non-angular coordinates are $-\infty <t<\infty$ and $r> 0$,
while for $k=-1$ we have 
$r > 1/\lambda$.

Due to Corollary 2 in \cite{signlong}, branes with a change of signature
require an asymmetric set-up, so that we need to glue a region of
AdS${}_{5}$ with a region of another anti-de Sitter space
$\widetilde{\mbox{AdS}}_5$ ---with a different $\tilde\Lambda_{5}$.
All quantities referring to
$\widetilde{\mbox{AdS}}_5$ will carry an
overtilde. The gluing is performed across appropriate hypersurfaces of
AdS$_5$ and $\widetilde{\mbox{AdS}}_5$ which are mutually identified,
thereby producing the brane/shell $\S$.

For simplicity 
we will only consider
branes $\S$ with spherical, plane or hyperboloidal symmetry, i.e.
a sym\-me\-try-preserving matching \cite{mps}.
Then, ignoring the angular coordinates, the
corresponding hypersurfaces are given in parametric form by
$\S : \{t=t(\xi),\,  r=r(\xi)\}$ and
$\tilde{\S} : \{ \Bt=\Bt(\xi),\,  \Br=\Br(\xi) \}$.
The matching implies \cite{signlong} that the first fundamental form on
the brane reads
\begin{equation}
\left.{d s^2}\right|_{\S}= \ff(\xi) d\xi^2 + a^2(\xi) 
d \Omega_{k}^2
\label{eq:ds2FLRW_N}
\end{equation}
where $a(\xi)$ is defined by
$r(\xi) = \Br(\xi)\equiv a(\xi) $
and $N(\xi)$ 
controls the embedding functions $t(\xi)$ and $\Bt(\xi)$ via
\begin{equation}
    \dot{t} = \frac{\sigma a}{k + \lambda^2 a^2} \sqrt{ \frac{\dot{a}^2}{a^2}
      - \ff \left ( \frac{k}{a^2} + \lambda^2 \right) } \label{eq:ts}
\end{equation}
and a similar equation for $\tilde t$ in terms of $\Blambda$, 
where $\sigma$ (and $\tilde\sigma$) are two signs and the dot stands for
$d/d\xi$. $N(\xi)$ and $a(\xi)$ are arbitrary functions 
only restricted to satisfy that
both square roots, in (\ref{eq:ts}) and its tilded version, are real.

 From (\ref{eq:ds2FLRW_N}), the brane $\S$ 
changes
signature if $N$ changes sign. The changes of signature
happen at ``instants'' 
$\xi= \xi_s$ 
where $\ff$ becomes, or stops being, zero. 
The set $S=\{\xi_s\}$ of all such points is called the
{\em signature-changing set}  of $\S$. In general, the brane has a
Lorentzian phase $\S_L$ where $N<0$, an Euclidean phase $\S_E$ defined
by $N>0$, and a null phase $\S_0$ 
where $N=0$. 

The Lorentzian part $\S_L$ describes a Robertson-Walker spacetime
with $\xi$ related to the standard cosmic time $T(\xi)$ by
\begin{equation}
\dot{T} = \sqrt{-\ff} \hspace{3mm} \mbox{on $\S_L$.}\label{cosmic}
\end{equation}
From the point of view of this Lorentzian phase, the Lorentzian
geometry becomes singular at $S\cap \overline{\S_{L}}$.  We want to
describe the type of singularity that any observers living on $\S_L$
will {\em believe to see} there. Before that, however, let us remark
that this `singularity' concerns {\em only} the ``Lorentzianity'' of
the brane, and can {\it only} be conceived from the {\em inner point
of view} of the Lorentzian phase $\S_L$. {\em Both the bulks and the
brane $\S$ are totally regular everywhere} for regular functions
$N(\xi)$ and $a(\xi)>0$.

In order to ascertain how the scientists living within $\S_{L}$
interpret the `singularity' at $S\cap \overline{\S_{L}}$, one first
needs to compute the energy-momentum tensor $\ttt_{\mu\nu}$ on the
brane. To this end 
one must use 
the appropriate {\em generalized Israel formula}, which
was presented in \cite{MarsSenovilla93} and applied in \cite{signlong} for 
the case under consideration.
$\ttt_{\mu\nu}$
has a simple eigenvalue $\hat\varrho$, a triple one $\hat p$
associated to $d\Omega_{k}^2$, and the fifth vanishes.
$\hat\varrho$ reads explicitly \cite{letter,signlong}
\begin{eqnarray}
&&\frac{\kappa^2_5}{3} \epsilon_1\sigma
\hat\varrho \left(2 \frac{\dot a^2}{k+a^2\lambda^2}-\ff\right)=\nonumber\\
&&
\sqrt{\frac{\dot{a}^2}{a^2}-\ff \left (\frac{k}{a^2}+\Blambda^2 \right)}
- \sqrt{\frac{\dot{a}^2}{a^2}- \ff \left( \frac{k}{a^2}+\lambda^2
\right)},\label{nova}
\end{eqnarray}
where $\kappa_{5}^2$ is the 5-dimensional gravitational coupling
constant and $\epsilon_{1}$ is a sign selecting which region bounded
by $\S$ in AdS$_{5}$ is to be matched with which region bounded by
$\tilde\S$ in $\widetilde{\mbox{AdS}}_5$, see
\cite{FST,signlong}.
$\hat\varrho$ and $\hat p$ are related by
\begin{equation}
\dot{\hat\varrho}+
\frac{d}{d\xi}
\left(
\log\frac{|2 \frac{\dot a^2}{k+a^2\lambda^2}-\ff|}
{\sqrt{|N|}}
\right) \hat\varrho+
3\frac{\dot{a}}{a}(\hat\varrho+\hat p)=0,
\label{eq:pre-cons}
\end{equation}
which has the interpretation of a continuity equation. For
signature-changing branes, $\ttt_{\mu\nu}$ is affected \cite{letter,signlong}
by a
normalisation freedom related to the choice of volume element
on the brane. Irrespective of this,
one can check that 
$\ttt_{\mu\nu}$, 
$\hat\varrho$ and $\hat p$ are regular everywhere
for regular 
$\S$. No singularitities arise at the signature change or elsewhere.

The question arises of how the observers living within the 
Lorentzian phase of the brane may interpret these facts, and the
observations which they perform, 
if they believe (erroneously) that their universe is Lorentzian
everywhere. It turns out that there are two different possibilities
according to their level of misinformation.

The first possibility arises if the scientists living in $\S_{L}$ {\em
know} that the bulk universe is 
5-dimensional and they live on
a 4-dimensional braneworld, but they do not consider
signature changes as feasible. They will assume 
$\ff < 0$ 
everywhere, and choose the cosmic time $T$ of (\ref{cosmic}) to describe
the age of the universe, hence from
(\ref{eq:ds2FLRW_N})
\begin{equation}
\left.{d s^2}\right|_{\S_L}= - d T^2 +
a^2 d \Omega_{k}^2 \, .\label{oldRW}
\end{equation}
They will also naturally normalize $\hat\varrho$
and $\hat p$ according to 
\begin{equation}
{\varrho} \equiv
\hat \varrho~ \frac{|2 \frac{\dot a^2}{k+a^2\lambda^2}-\ff|}{\sqrt{|\ff|}},~
{p}\equiv
\hat p ~ \frac{|2 \frac{\dot a^2}{k+a^2\lambda^2}-\ff|}{\sqrt{|\ff|}},\label{1/N}
\end{equation}
so that the conservation law (\ref{eq:pre-cons}) adopts
its standard form
\begin{equation}
\dot{{\varrho}}+3\frac{\dot{a}}{a}({\varrho}+{p})=0.
\label{eq:cons}
\end{equation}
This normalization 
corresponds to the canonical 
Robertson-Walker volume element.
Therefore, ${\varrho}$ and ${p}$ 
are \emph{the energy
density and pressure measured within 
$\S_L$.}

However, things behave quite differently for signature-changing
branes in comparison with purely Lorentzian ones 
regarding the `end of time' (`beginning'
for signature changes in the past).
In signature-changing branes these
finales can have a {\em non-vanishing $a$}, with finite $\dot a$ and $\ddot
a$, 
but where $a'= da/dT$ diverges.
Indeed, $a' =\dot{a}/\sqrt{-N}$ {\em diverges
necessarily} when approaching the signature-changing set
$S\cap\overline{\S_L}$ given that
$\dot a|_S\neq 0$ (otherwise,
since $N=0$ on $S$, (\ref{eq:ts}) 
would imply $\dot t|_S=\dot{\Bt}|_S=0$,
which is impossible \cite{signlong}). Thence, the Hubble function
$H\equiv a'/a$ diverges necessarily at $S\cap\overline{\S_L}$.
This behaviour cannot be found in pure Lorentzian brane cosmologies.

Observe that (\ref{1/N}) may seem to imply that ${\varrho}$ and ${p}$
diverge when approaching 
$S\cap \overline{\Sigma_{L}}$, where $N\rightarrow 0$. Actually, this is not
the case because one can prove that ${\varrho}$ {\em vanishes} at the
signature change,
and that $p$ can also be regular there in many cases
\cite{signlong}.
To see this, let us rewrite (\ref{nova}) in terms of the physical quantity $\varrho$ 
\begin{equation}
\frac{\kappa^2_5}{3}\, \varrho=
\sigma\epsilon_1
\left(
\sqrt{\frac{{a'}^2+k}{a^2}
+\Blambda^2 }
- \sqrt{\frac{{a'}^2+k}{a^2}
+\lambda^2}\,
\right),
\label{eq:Friedmann_L}
\end{equation}
which, we remark, holds only on $\S_L$. From (\ref{eq:Friedmann_L}) 
it is easy to
find \cite{signlong} the following {\em bounds} for the
energy density:
$\frac{\kappa_5^2}{3} |\varrho| \leq  \sqrt{|\Blambda^2-\lambda^2|} \, .$
This inequality is strict for $k=1$, while for $k=0$ we have the stronger 
$\frac{\kappa_5^2}{3} |\varrho| \leq  |\Blambda-\lambda|$.
Now, (\ref{eq:Friedmann_L}) 
implies also that $\varrho$ {\em vanishes at the change of signature},
\begin{equation}
\lim_{x\rightarrow S\cap\overline{\S_L}} \varrho =\lim_{x\rightarrow S\cap\overline{\S_L}} \frac{3\sigma\epsilon_1}{2\kappa^2_5}
  \frac{\Blambda^2-\lambda^2}{|H|}=0.
\label{eq:lim_rho}
\end{equation}
This limit together with (\ref{eq:cons}) leads to
\begin{eqnarray*}
\lim_{x\rightarrow S\cap\overline{\S_L}}   p  =
\lim_{x\rightarrow S\cap\overline{\S_L}}  \frac{\sigma\epsilon_1}{4\kappa^2_5}
  (\Blambda^2-\lambda^2) 
\frac{a^2}{\dot a |\dot a|} \frac{\dot{N}}{\sqrt{-N}}
\end{eqnarray*}
which shows that $p$ can be regular if $\dot N|_{S}= 0$. 
Therefore, by choosing appropriately the hypersurfaces in AdS$_{5}$
and $\widetilde{\mbox{AdS}}_5$ {\em one can construct 
signature-changing branes with both $\varrho$ and $p$ finite and
well-behaved everywhere on $\overline{\S_L}$}. As a matter of fact,
signature-changing branes with
$\ttt_{\mu\nu}$ satisfying some
desirable energy conditions can be built, as we will show below.


Going to a more severe level of misinformation of the
scientists living within $\S_{L}$, assume not only that
they are uninformed about 
the possibility of signature changes, but also 
that they even {\em ignore} they live in a
brane of a 5-dimensional bulk. They will again assume that $\ff < 0$
holds everywhere, and choose the cosmic time $T$ to
describe the age of the universe so that the line-element is given
by (\ref{oldRW}). In  addition, they will also use their
favourite gravitational theory to describe the Universe and to compute
its energy-momentum content. This will generally lead to an
`energy-momentum tensor' which has nothing to do with the
genuine $\ttt_{\mu\nu}$ on the brane.

If they 
consider
General Relativity (GR)
as the correct gravitational theory,
they will 
compute the eigenvalues of the {\em Einstein tensor} 
of the line-element (\ref{oldRW})
on $\S_L$
\begin{eqnarray*}
8\pi G\, \varrho^{(GR)}+\lambdafour&=&\frac{3}{a^2}(a'^2+k),
\\
8\pi G \, p^{(GR)}-\lambdafour&=&-2\frac{a''}{a}-\frac{1}{a^2}(a'^2+k),
\end{eqnarray*}
which obviously diverge at the `singularity' placed on the
signature-changing set $S\cap \overline{\S_L}$.  Here, $\lambdafour$
is the GR cosmological constant. Observe that this `singularity' would
be interpreted as a {\em sudden singularity} in the sense of \cite{CV}
(generalising those in \cite{B}),
as the scale factor $a$ remains finite but $a'$ becomes unbounded.
They have also been termed as `type III' singularities
\cite{NOT} and `big freeze' \cite{bigfreeze}, see also \cite{FL}.
These singularities
require the violation of the energy conditions \cite{CV}, but of
course {\em this refers} to the energy conditions satisfied by $
\varrho^{(GR)}$ and $p^{(GR)}$. Hence, this violation, or equivalently
the existence of phantom or dark energy components, may be just an
illusion caused by a forthcoming change of signature in the brane.

The relationship between $\varrho^{(GR)}$
and the true $\varrho$ 
on the brane
can be deduced from (\ref{eq:Friedmann_L}), and is given by \cite{signlong}
\begin{eqnarray*}
&&\frac{\kappa^2_5}{3}\varrho=\sigma\epsilon_1
\left[
\sqrt{\frac{8\pi G}{3}\varrho^{(GR)}+\frac{\lambdafour}{3}+\Blambda^2 }
\right.\\
&&\left.
-\sqrt{\frac{8\pi G}{3}\varrho^{(GR)}+\frac{\lambdafour}{3}+\lambda^2 }
\right],\\
&&8\pi G\,  \varrho^{(GR)}+\lambdafour+3\Blambda^2 =
\frac{27}{4\kappa_5^4\varrho^2}\left(\Blambda^2-\lambda^2+\frac{\kappa_5^4}{9}\varrho^2\right)^2.
\end{eqnarray*}
It is interesting to see that, in this setting, the
GR `singularity' $\varrho^{(GR)}\rightarrow \infty$ is
manifestly due to the {\em vanishing} of the proper energy density at
the change of signature.


Let us show with an explicit example how a seemingly sudden singularity
can be described with a regular signature changing brane
satisfying the strong energy condition.
For concreteness let us look for an equation of state of the form
$p=C \varrho^{\alpha}$ with $C>0$. To restrict the possible values
of $\alpha$ 
we 
need to study the behaviour 
$\varrho$ and $p$ near the signature change.

Let us assume that $N(\xi)$ approaches zero at the signature
change located at $\xi = \xi_f$ as 
$N=(\xi-\xi_f)^m M(\xi)$ where $m$ is an odd
integer and $M(\xi)$ is a regular function, positive at $\xi_f$.
Since $\dot a (\xi_f) \neq 0$ 
the limit (\ref{eq:lim_rho}) ---coming from
the Lorentzian part ($\xi<\xi_f$)---
reduces to $\varrho \sim 
(\xi_f-\xi)^{m/2}$ 
where $\sim$ means that both terms are equivalent infinitesimals.
In order to get a regular $p$ we need $m\geq 3$, so that $\dot N (\xi_f)=0$.
On the other hand, the conservation law (\ref{eq:cons})
leads to $p \sim (\xi_f-\xi)^{(m-2)/2}$, which implies
$p \sim \varrho^{(m-2)/m}$  near $\xi_f$.
Thus, if we want to keep a power law equation
of state all along $\S$,
we must set
$p=C \varrho^{(m-2)/m}$.
The conservation law (\ref{eq:cons}) integrates to 
$$
\varrho=C^{m/2}\left(
\aa^{-6/m}-1\right)^{m/2},
$$
where $\aa\equiv a/a_S$ and the integration constant
has been chosen as $a_S = a(\xi_f)$
in order to enable a change of signature.
%
Inserting this into (\ref{eq:Friedmann_L}) we get,
after some algebra,
%
\begin{eqnarray}
  \label{eq:a_quad}
  \dot \aa^2 &=&\frac{M(\xi_f-\xi)^m}{4\beta \aa^4 (1-\aa^{\frac{6}{m}})^{m}} F(\aa),
\end{eqnarray}
where $\beta\equiv \kappa_5^4 C^m/9$ and 
\begin{eqnarray*}
\label{eq:defF}
&&F(\aa)\equiv
\left[\aa^6 (\Blambda-\lambda)^2 -\beta (1-\aa^{\frac{6}{m}})^{m}\right]\\
&&\times\left[\aa^6 (\Blambda+\lambda)^2 -\beta (1-\aa^{\frac{6}{m}})^{m}\right]
-4\beta \frac{k}{a_S^2}\aa^4(1-\aa^{\frac{6}{m}})^{m}.
\end{eqnarray*}
Solutions of (\ref{eq:a_quad}) can be proven
to satisfy that the square roots in (\ref{eq:ts}) and its tilded counterpart
are real.

Eq.(\ref{eq:a_quad}) being quadratic,
it contains two branches. 
To choose the proper one,
note that 
$\aa(\xi_f)=1$ by construction and $F(1)=(\Blambda^2-\lambda^2)^2 >0$. 
Furthermore, at points of $\S_L$ near $S$, where 
$\xi<\xi_f$, 
we need $\aa<1$ to keep a well-defined $\varrho$.
Hence $\dot \aa >0$ 
near $S$ and the
positive square root in (\ref{eq:a_quad})
has to be chosen.
The solution for $\aa(\xi)$ depends strongly on 
$F(\aa)$, and in particular on its zeros
and their order. For $k=0,1$ the existence of two simple zeroes
of $F(\aa)$ in $\aa\in(0,1)$ is ensured. 
Let us focus for simplicity on the case $k=1$, and
let $\aa_0\in (0,1)$ denote the zero of $F(\aa)$
closest to $\aa=1$. We must ascertain if
a solution exists extending from $\aa=\aa_0$ to  $\aa=1$,
i.e. that for a finite $\xi_0$, $\aa(\xi_0)=\aa_0$ holds.
Integrating (\ref{eq:a_quad}), this
amounts to showing that, for some finite $\xi_0$,
\[
4\beta \int^1_{\aa_0} \frac{\aa^2 (1-\aa^{\frac{6}{m}})^{m/2}}{\sqrt{F(\aa)}} d\aa=
\int^{\xi_f}_{\xi_0} M(\xi) (\xi_f-\xi)^m d\xi,
\] 
with both integrals convergent. Since
$F\sim (\aa-\aa_0)$ near $\aa_{0}$, the integral on the left 
trivially converges.
The integral on the right
converges for any finite $\xi_0$, provided $M(\xi)$ stays bounded.
Thus, for a large class of positive functions $M(\xi)$ (in particular
for all those bounded away from zero) a finite
$\xi_0$ fulfilling the equality does exist. 

Let us assume that one such 
$M(\xi)$ has been chosen. The corresponding solution $\aa(\xi)$
satisfies $\dot\aa(\xi_0)=0$ by construction. It is easy to check 
that $\ddot\aa(\xi_0)>0$, which implies that the negative branch
of (\ref{eq:a_quad}) must be taken for $\xi<\xi_0$ (near $\xi_0$).
Using a similar argument as before, the solution $\aa(\xi)$
increases (as $\xi$ decreases) until reaching again
$\aa=1$ at some finite $\xi_b <\xi_0<\xi_f$.
In principle, $\dot\aa$ might diverge there, leading to a singular
brane. This can be avoided if and only if $M=(\xi_b-\xi)^m P(\xi)$
for some regular function $P(\xi)$ positive in $[\xi_b,\xi_f]$.
The simplest way to ensure this behaviour is to choose
$M(\xi)$ even with respect to $\xi= \xi_0$. This implies the same even property for $\aa(\xi)$.
It should be stressed, however, that this is just
one among many possibilities.



Let us stress that the explicit Lorentzian cosmological model,
described by $a(T)$, depends on $N(\xi)$ only through its bevahiour
near its zeros. The free function $P$ is irrelevant as it can always
be reabsorbed with a coordinate change within the Lorentzian
phase. This is already apparent in (\ref{eq:a_quad}), from which a simple
calculation 
gives
\begin{equation}
  \label{eq:app_pre}
  a''= a_S\aa''=(-1)^{m+1}\frac{a_S}{2}\frac{d}{d\aa}
  \left(\frac{F(\aa)}{4\beta \aa^4 (1-\aa^{\frac{6}{m}})^{m}}\right)
\end{equation}
where the dependence on $P$ has vanished.

\begin{figure}
  \includegraphics[width=6cm]{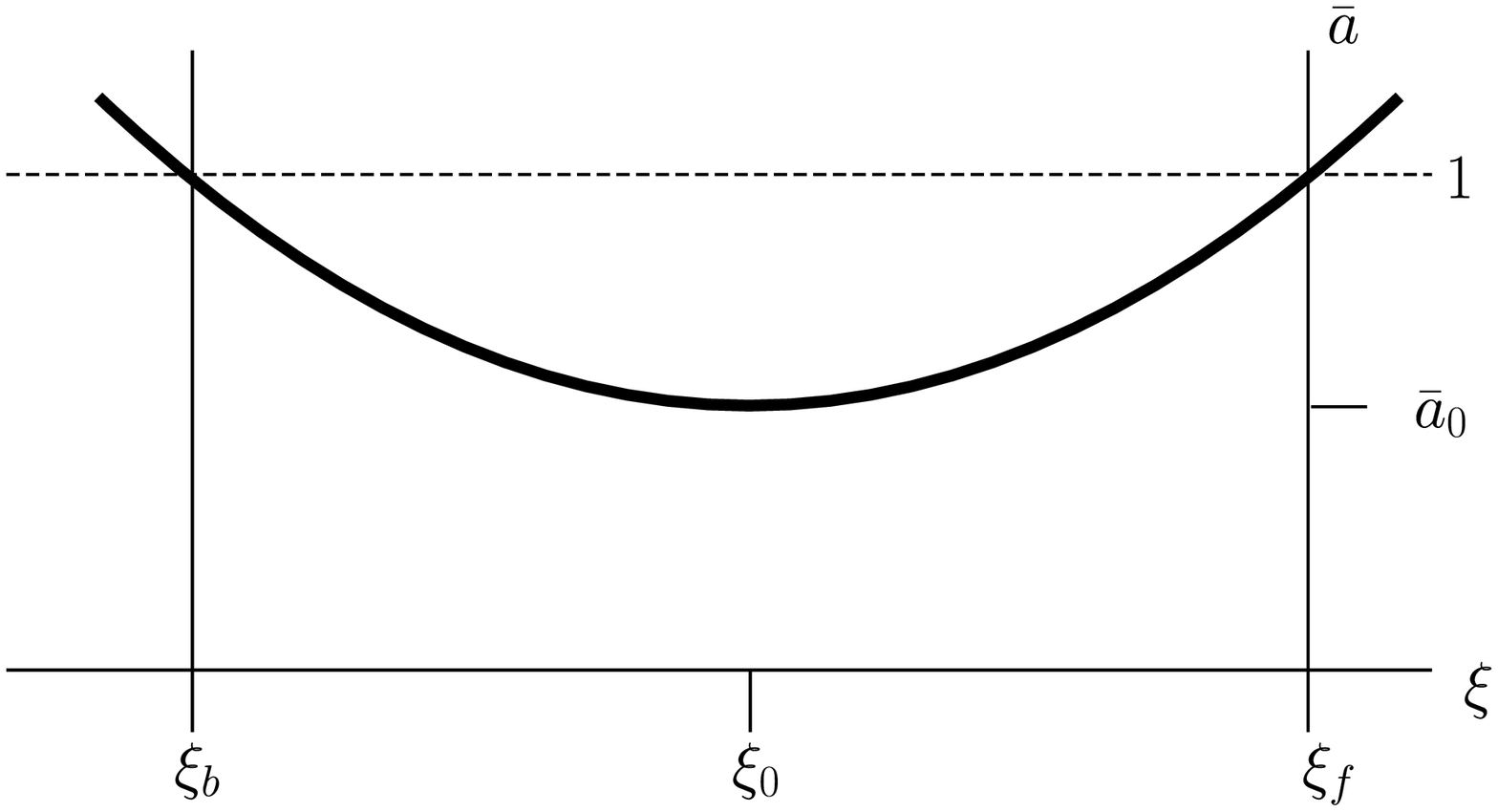}\\
  \parbox[c]{4.2cm}{\includegraphics[width=4.2cm]{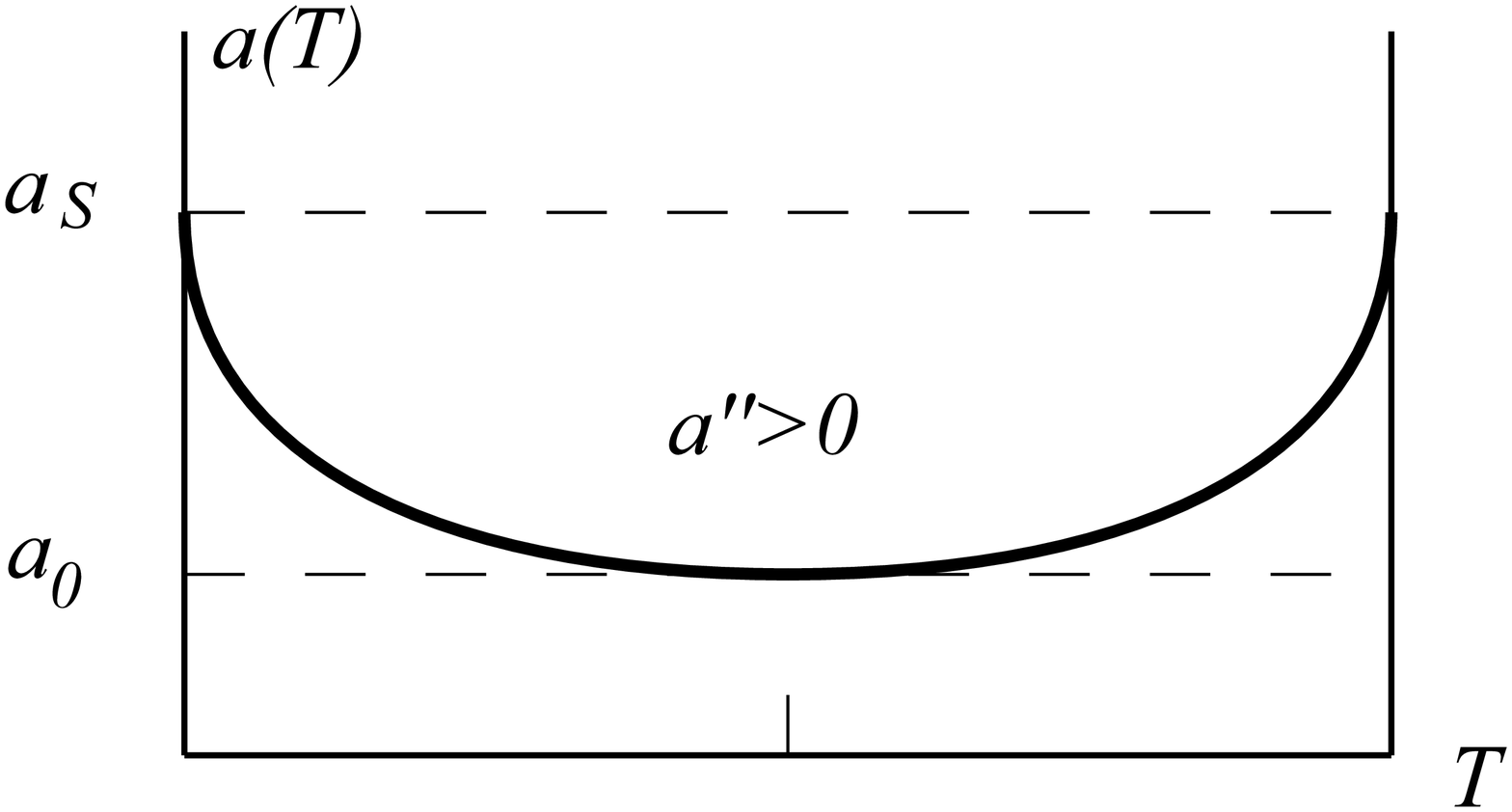}}%
  \hfill%
  \parbox[c]{4.2cm}{\includegraphics[width=4cm]{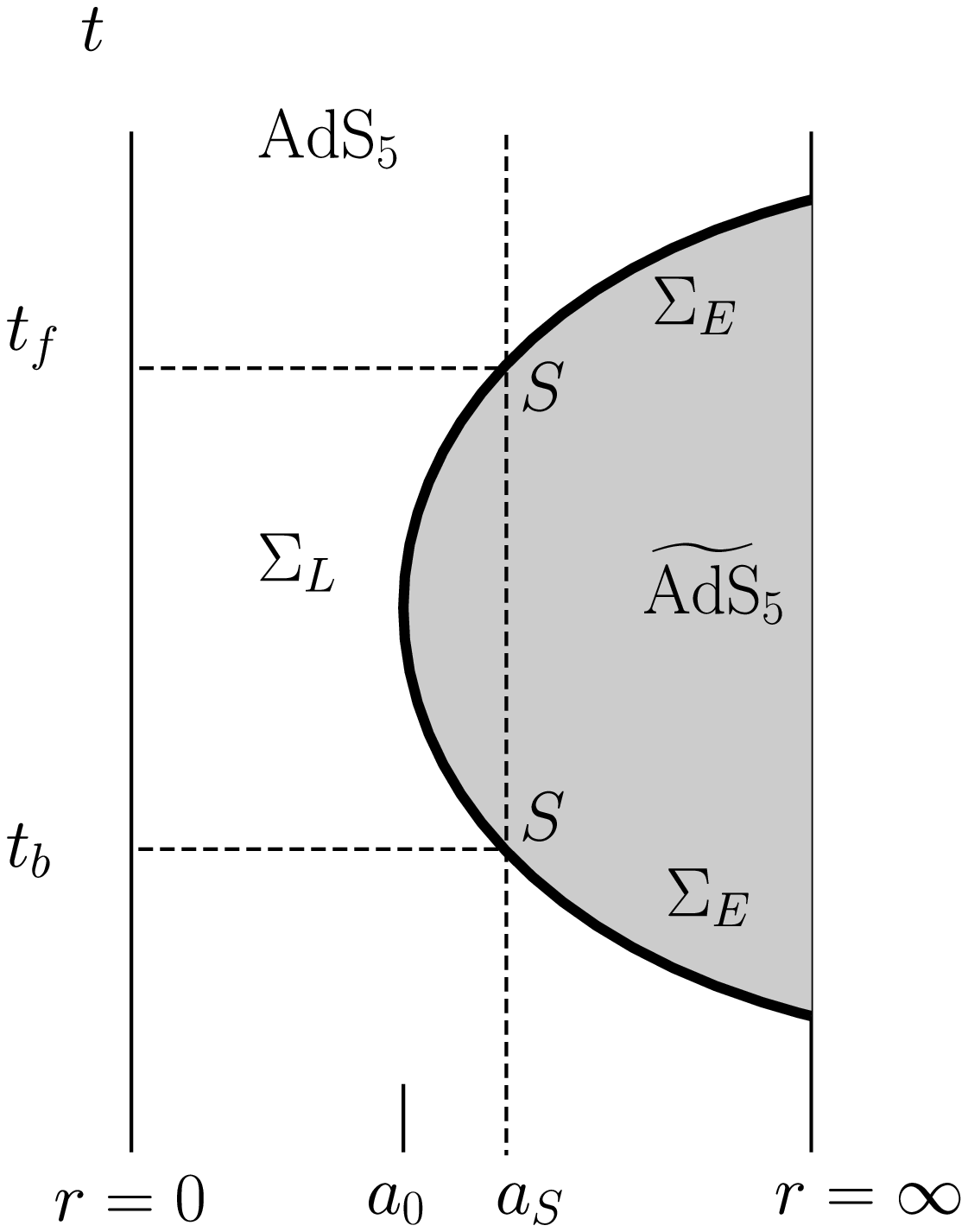}}
\caption{The Lorentzian
phase starts with a signature change at $\xi_b$, with
a scale factor $a=a_S$ ($\aa=1$) and
vanishing density $\varrho$ and pressure $p$. The model
contracts until it reaches the minimum scale $a_0$ where
$\varrho$ 
attains its maximum, which represents a {\em little bang} (regular bounce) \cite{S}.
The model then `bounces' to an expanding era with accelerated expansion
($a''>0$), in which both $\varrho$ and $p$
decrease, until another change of signature occurs at
$a = a_S$, with $\varrho |_{S}=p|_{S}=0$. The density and pressure
remain non-negative 
everywhere on the Lorentzian phase. The picture on the
bottom right
describes the whole brane (with its Lorentzian, Euclidean and null phases)
embedded in the AdS$_5$ bulks, with $\sigma=-\epsilon_1=1$
and $\Blambda<\lambda$, so that $\varrho> 0$ in $\S_L$.
Observe that $\varrho|_{\S_{L}}> 0$ requires
$\Lambda_{5}<\widetilde{\Lambda_{5}}<0$ so that the `eternal' AdS$_{5}$
is more stable. This is very reasonable physically \cite{letter}.
}
\label{fig:modelet}
\end{figure}


Regarding 
the behaviour of 
the cosmic acceleration at the signature change,
we need to evaluate $\aa''$ in the limit $\aa=1$. The calculation gives
%
\[
\lim_{\aa\to 1}a''=
\lim_{\aa\to 1} \frac{2 a_S 
  (\Blambda^2-\lambda^2)^2}{\beta (1-\aa^{\frac{6}{m}})^{m+1}}=+\infty \, .
\]
Thus $a''(T)$ must be positive 
near 
the signature changes, and 
\emph{the seemingly `sudden singularity' (of big-freeze type)
in the future 
is approached 
while in an increasingly accelerated expansion epoch}
(see Fig. \ref{fig:modelet}).

We have that $a''(T)$ must be positive both around $\xi_0$ and $\xi_f$, that is
(i) at the beginning of the expansion epoch (when the energy
density attains its maximum), which can account
for an inflation era, and (ii) in the accelerated expansion
epoch previous to the final `big freeze'. On the other hand, the very particular class of models with $k=1$ and equation of state $p=C \varrho^{\alpha}$ that we have built cannot present an epoch of decelerated expansion with
$a''(T)<0$ ---in order to stop the initial inflation---,
as it can be proven by using (\ref{eq:app_pre}).
However, due to the large freedom
in the equation of state it seems likely that 
many models will exist with $a''\leq 0$ for some period
after the initial inflation epoch.

We thank L. Fern\'andez-Jambrina and R. Lazkoz
for their suggestions, and also M. Bouhmadi-L\'opez for her comments.
MM was supported by the projects
FIS2006-05319 of the Spanish MEC and
SA010CO5 of the Junta de Castilla y Le\'on. 
JMMS thanks support under grants FIS2004-01626 (MEC) and GIU06/37 of the University of the Basque Country (UPV).
RV is funded by the Basque Government Ref. BFI05.335 and
thanks support from project GIU06/37 (UPV).

\end{document}